\documentclass[11pt]{article}
\usepackage[a4paper, left=2cm, right=2cm, top=2cm, bottom=2.5cm]{geometry}

\usepackage[T1]{fontenc}
\usepackage{graphicx}
\usepackage{subcaption}
\usepackage{calc}
\usepackage{amstext}
\usepackage{amsmath}
\usepackage{amssymb}
\usepackage{amsthm}
\usepackage{algorithm2e}
\usepackage{braket}
\usepackage{url}
\usepackage{hyperref}
\usepackage{booktabs}

\newcommand {\R}{\mathbb{R}}

\newcommand{\keywords}[1]{\noindent\textbf{Keywords:} #1\vspace{1em}}

\title{Quantum Optimization-Based Route Compression for Efficient Navigation Systems}

\author{
    Shunsuke Sotobayashi\\
    blueqat Research, Minato, Tokyo, Japan\\
    \texttt{derwind0707@gmail.com}
    \and
    Yuichiro Minato\\
    blueqat, Minato, Tokyo, Japan\\
    \texttt{minato@blueqat.com}
    \and
    Takao Tomono\\
    Keio University Sustainable Quantum AI Center\\
    \small Keio University, Graduate School of Science and Technology, Yokohama, Kanagawa\\
    \small Graduate School of Media and Governance, Fujisawa, Kanagawa\\
    \small Keio University Sustainable Quantum Artificial Intelligence Center (KSQAIC), Minato, Tokyo, Japan\\
    \texttt{takao.tomono@ieee.org}
}

\date{}

\begin{document}

\maketitle

\begin{abstract}
We present a novel quantum optimization-based route compression technique that significantly reduces storage requirements compared to conventional methods. Route optimization systems face critical challenges in efficiently storing selected routes, particularly under memory constraints. Our proposed method enhances route information compression rates by leveraging Higher-Order Binary Optimization (HOBO), an extended formulation of Quadratic Unconstrained Binary Optimization (QUBO) commonly employed in quantum approximate optimization algorithms (QAOA) for combinatorial optimization problems. We implemented HOBO on real-world map data and conducted comparative analysis between the traditional Ramer--Douglas--Peucker (RDP) algorithm and our proposed method. Results demonstrate that our approach successfully identifies yielding improved compression efficiency that scales with data size from candidate routes. Experimental validation confirms the technique's viability for practical applications in navigation systems where memory constraints are critical. The HOBO formulation allows for representation of complex route that would be difficult to capture using classical compression algorithms. Our implementation demonstrates up to 30\% improvement in compression rates while maintaining route fidelity within acceptable navigation parameters. This approach opens new possibilities for implementing quantum-inspired optimization in transportation systems, potentially providing more efficient navigation services. This work represents a significant advancement in applying quantum optimization principles to practical transportation challenges.
\end{abstract}


\maketitle

\section{Introduction} \label{sec:introduction}

While much of the research surrounding quantum computers is focused on hardware-related topics such as those with superconductors, neutral atoms, and semiconductors, there have also been many software-related proposals for quantum algorithms \cite{Kim2023,Evered2023,Neyens2024}. On the other hand, there have been few proposals related to memory. Algorithms and memory are very important fundamental technologies in computers.

Quantum computers are expected for security problems, quantum simulation, machine learning, and mathematical optimization.

Here we would like to give a brief overview of machine learning, both classical and quantum. Classical machine learning has received increasing attention since AlexNet~\cite{Krizhevsky_Sutskever_Hinton2012}, which uses deep learning convolutional neural networks, won first place at ILSVRC 2012, and has been followed by a series of successful image classification models such as VGG16~\cite{Simonyan2015} and ResNet~\cite{He_2016_CVPR}. Other applications include data generation. Examples include Variational Autoencoders (VAEs)~\cite{Kingma2014} and Generative Adversarial Networks (GANs)~\cite{Goodfellow2014,Zhu_2017_ICCV,karras2018progressive}, which utilize game theory. All of this is implemented as a mathematical model with many parameters. Each task is performed by automatically adjusting these parameters to optimal values. On the other hand, there are problems behind the success of classical machine learning. Deep learning consumes a large amount of power as it drives a large number of GPUs, and tends to produce high CO2 emissions. From this perspective, there is growing interest in machine learning using quantum computers, which are expected to operate with lower power and CO2 emissions. Machine learning using quantum computers has been studied for applications to image classification models~\cite{Havlicek2019,Cong2019}, data generation~\cite{Zoufal2019}, and language models~\cite{Basile2017,Bausch2020}, as well as classical machine learning. Apart from these, applications of machine learning using quantum computers include applications in material science and quantum chemistry, and combinatorial optimization problems. In the material science and quantum chemistry application, an algorithm called VQE is used to solve the eigenvalue problem of the Schr\"{o}dinger equation using variational methods for the properties of molecules and materials. In combinatorial optimization problems, an algorithm called QAOA is used to find the best combination from an extremely large number of candidates. Solving the combinatorial optimization problem is very promising.

In fact, combinatorial optimization problems are often used to solve many of these social problems. Examples include: car navigation system route finding---the problem of finding the shortest way between two points; delivery planning---the problem of finding a route to deliver parcels to everyone in the shortest time; investment---the problem of maximizing profits with a certain amount of money; and work schedule---the problem of finding a work schedule that satisfies all employees' working hours and other preferences.

In traffic flow problems, quantum annealing techniques can be used for the purpose of finding the optimal combination of paths among many path combinations. In quantum annealing, the combinatorial optimization problem is rewritten in the form of an energy function (Ising Hamiltonian) based on the Ising model, and a quantum annealing machine is used to find the minimum energy by converging to the ground state based on the quantum adiabatic theorem~\cite{Kato1950}. The state of the system that achieves the minimum energy, when properly formulated, corresponds to the optimal solution of a combinatorial optimization problem. An implementation of an algorithm based on a principle similar to this technique on a gated quantum computer is called the Quantum Adiabatic Algorithm (QAA). QAA has challenges in terms of scalability for large-scale problems and error tolerance for gated quantum computers. In contrast, the Quantum Approximate Optimization Algorithm (QAOA) has been proposed as a method to reduce the principle of the quantum adiabatic algorithm to discrete gate operations and to obtain an approximate optimal solution.

Besides solving combinatorial optimization problems, another very important issue is to keep a record in memory. Some familiar examples are the problem of storage and network transfer of the ever-exploding text, image, and voice data, which are detailed in \cite{Fitriya_Purboyo_Prasasti2017,JAYASANKAR2021}. In the area of transportation, a variety of transportation methods continue to be developed, and route information is handled in a variety of devices, including car navigation systems, smartphone applications, and GPU smartwatches. Not only text, images and audio, but also route information is desired to be smaller in terms of storage in devices with limited storage capacity, or in terms of network bandwidth and transfer rate when sending data to a server. For this purpose, it is necessary to compress the data efficiently and reduce the size of the data before saving it.

One of the classical solutions is the well-known Ramer--Douglas--Peucker (RDP) algorithm \cite{RAMER1972244,Douglas_Peucker_1973,Visvalingam_Whyatt_1993,Bo_et_al_2020}.

RDP algorithm can provide effective compression for relatively simple structures. However, it has been pointed out that RDP is a point-based algorithm and is not suitable for generalization of complex lines, especially those that retain map features \cite{Visvalingam_Whyatt_1993}. It has also been pointed out that simplifying linear objects by applying RDP to them tends to produce results that contain self-intersection errors \cite{Bo_et_al_2020}.

Regarding the performance of RDP, it is pointed out that the worst-case performance of \cite{Hershberger_Snoeyink_1992} is $\mathcal{O} (n^2)$ for the number of points $n$. In their work, \textit{the path hull algorithm} that achieves $\mathcal{O} (n \log_2 n)$ in computational complexity is proposed.

Data compression in RDP is good enough, however, reliance on a single hyperparameter for global control can lead to compression results that are not always optimal. In contrast, it is desirable to incorporate local control. In other words, multiple local path candidates are considered and the optimal path that satisfies the conditions is selected from among them. Such a problem is generally called a ``combinatorial optimization problem,'' and is known as one of the most promising applications of quantum computers because it can be solved with a small amount of computation by using the principles of quantum mechanics. In this study, we propose a quantum optimization-based route compression technique and our method uses HOBO (Higher-Order Binary Optimization), which is an extension of QUBO (Quadratic Unconstrained Binary Optimization), a conventionally popular formulation in QAOA, for solving ``combinatorial optimization problems''. In Sec.~\ref{sec:related_works}, we introduce RDP and QUBO. In Sec.~\ref{sec:proposed_method}, we explain our proposal with mathematical formulas, and in Sec.~\ref{sec:experiment}, we show the results of applying our proposal to three actual cases (a route from Beppu Station to Yufuin, a route of Okuhida Trail Run Course and a route of Iroha-zaka Road). In Sec.~\ref{sec:discussion}, we describe the optimization issues and current limitations, and in Sec.~\ref{sec:conclusion}, we summarize the practical aspects of the proposal and discuss future prospects.

\section{Related works} \label{sec:related_works}

The Ramer--Douglas--Peucker algorithm \cite{RAMER1972244,Douglas_Peucker_1973} (RDP) is a well-known existing compression method for route data (see Alg.~\ref{alg:DouglasPeucker} in Sec.~\ref{sec:rdp_pseudo_code} for a pseudo code). RDP generates a simplified line from a given collection of points (polyline) by removing unnecessary points without significant loss of shape. This algorithm is used for data reduction, smoothing, and pattern recognition in maps and graphics. In \cite{Douglas_Peucker_1973}, the algorithm for reducing the number of points needed to represent a digitized line or its approximation is explained, quantifying lines obtained from maps and photographs as examples. Prior to the RDP algorithm, an existing method was known to set up a grid on a map and record Yes/No information on whether the grid and the route intersect. This method is time consuming and has the drawback that the grid size is fixed at the time of manufacture. Other methods also mechanically take each $n$ th point, but this can lead to over-compression that ignores the geometry, as described in \cite{Visvalingam_Whyatt_1993}, for example. This work focuses on changes in the area of figures composed of lines as points are reduced and addresses simplification of lines and point reduction based on the concept of ``effective area.'' They also point out that the RDP algorithm is a point-based algorithm, which is suitable for minimal simplification, but not for generalization of complex lines, especially generalization that preserves map features. While acknowledging the rationale behind the RDP's compression method, which assumes that ``the further point from any arbitrary anchor-floater line must be a critical point,'' they also question this assumption. A several recent studies are \cite{Bo_et_al_2020}. In \cite{Bo_et_al_2020}, they point out that simplifying linear objects by applying RDP to them tends to produce results that contain self intersection errors. To overcome this problem, a new vector line simplification algorithm based on the RDP algorithm but with monotonic chains and dichotomy is proposed and claimed to be superior in terms of efficiency and accuracy.

In quantum computation, an algorithm called QAOA (Quantum Approximate Optimization Algorithm)~\cite{Farhi_Goldstone_Gutmann_2014} is known, which is a hybrid algorithm combining conventional classical optimization algorithms and quantum mechanical principles. The algorithm is often applied to combinatorial optimization problems such as the max-cut problem (one of the problems in graph theory), the knapsack problem, and the traveling salesman problem, and takes advantage of the powerful computational capabilities of quantum computers. In our case, we want to exclusively select one of the paths which start at a node. For this purpose, it is useful to be able to handle, for example, the quantum state $\frac{1}{\sqrt{3}}(\ket{001} + \ket{010} + \ket{100})$ corresponding to selecting only one of the three candidates. For this purpose, it is known that good results can be achieved by making good use of a component called ``mixing Hamiltonian'' or ``mixer''. The mixer corresponding to a case like this is called a $XY$-mixer \cite{Hadfield_2019,Wang_2020}. A $XY$-mixer induces state transitions among candidates. In general, when using a mixing Hamiltonian, the corresponding quantum state must be prepared. For this $XY$-mixer, a simple method of preparation is known \cite{B_rtschi_2019}.

QAOA uses a specific type of quantum circuit to construct ansatz (Parameterized quantum circuit), but there is also a method using VQE (Variational Quantum Eigensolver)~\cite{Peruzzo2014} that can use more flexible ansatz. VQE is often used in quantum chemical calculations to find eigenvalues of the ground state of the Hamiltonian. In such cases, ansatz suitable for calculations such as UCCSD (Unitary Coupled Cluster Singles and Doubles) ansatz are often employed ~\cite{Sok20}, but beyond that, they are also applied to general combinatorial optimization problems. For example, in \cite{Matsuo2023}, a method similar to the use of $XY$-mixer in QAOA is employed, where the construction of the ansatz is devised to impose constraints on the solution candidates.

QUBO (Quadratic Unconstrained Binary Optimization) is a widely used formulation of expressing optimization problems in QAOA, and when used for path selection as in the present case, the aforementioned $XY$-mixer can be effectively utilized. In our study, we use HOBO (Higher-Order Binary Optimization), an extension of HOBO, to treat it as a higher-order optimization problem. The difference between QUBO and HOBO lies in the order of the interactions between variables. QUBO is, as the name implies, binary variable optimization, whereas HOBO is an extension of QUBO and represents higher-order unconstrained optimization problems. QUBO deals only with terms up to second order, while HOBO deals with terms of higher orders, i.e., third and higher. A study of QAOA using HOBO is known, for example, in \cite{Glos_Krawiec_Zimboras_2022}. In that work, using the Traveling Salesman Problem (TSP) as an example, they proposed a method to represent the problem with significantly fewer qubits than conventional encoding methods. On the other hand, although the number of qubits can be reduced, the number of terms increases exponentially, and thus the depth of the quantum circuit increases. This is discussed in Sec.~\ref{sec:discussion}.

The proposed method employs an efficient encoding scheme for the selected edges, but similar schemes are already known, for example, in \cite{Tan2021qubitefficient}, where it is called minimal encoding, and it is shown that it can encode with a logarithmic number of qubits for the problem size.

\section{Overview of RDP} \label{sec:overview_of_rdp}

Here we would like to give an overview of the RDP.

\begin{figure}[!ht]
\centering
\includegraphics[width=0.5\linewidth]{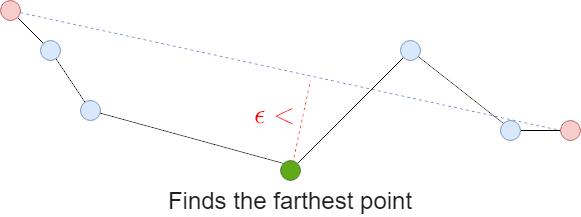}
\caption{RDP's first step. The furthest point C (green) from the line segment with the first point A (red) and the last point B (red) is marked and retained.}
\label{fig:rdp-1}
\end{figure}

\begin{figure}[!ht]
\centering
\includegraphics[width=0.5\linewidth]{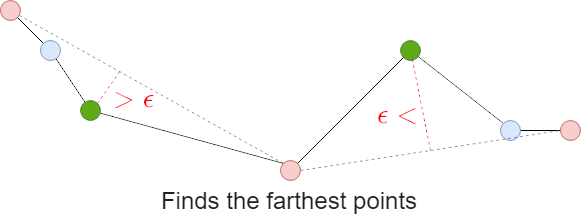}
\caption{RDP's second step. The furthest point D (green) from the line segment with the first point A (red) and the last point C (red), and the furthest point E (green) from the line segment with the first point C (red) and the last point B (red), are marked and retained.}
\label{fig:rdp-2}
\end{figure}

\begin{figure}[!ht]
\centering
\includegraphics[width=0.5\linewidth]{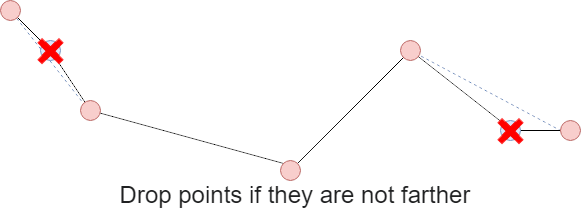}
\caption{RDP's subsequent steps. The same process continues to repeat itself as before, but if the furthest point (from the line segment) of interest falls below the threshold $\epsilon$, the point is not marked and discarded.}
\label{fig:rdp-3}
\end{figure}

\begin{figure}[!ht]
\centering
\includegraphics[width=0.5\linewidth]{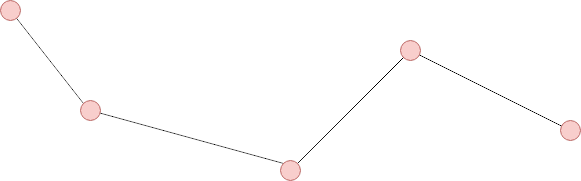}
\caption{RDP's final result. The process is complete when there are no more points to mark.}
\label{fig:rdp-4}
\end{figure}

The RDP algorithm divides the route recursively. First, the first point A and the last point B are automatically marked and retained. The furthest point C from the line segment with the first point A and the last point B as endpoints is found, which is also marked and retained (Fig.~\ref{fig:rdp-1}). It then recursively calls RDP algorithm itself using the first point A and the furthest point C (from the line segment), as well as point C and the last point B (Fig.~\ref{fig:rdp-2}). In this recursive call, if the furthest point (from the line segment) of interest falls below the threshold $\epsilon$, the point is not marked and discarded (Fig.~\ref{fig:rdp-3}). In this way, a compressed path is finally obtained as shown in Fig.~\ref{fig:rdp-4}.

For the pseudo code of the algorithm, please refer to Alg.~\ref{alg:DouglasPeucker} in Sec.~\ref{sec:rdp_pseudo_code}.

\section{Proposed method} \label{sec:proposed_method}

\subsection{Basic Idea}

\begin{figure}[!ht]
\centering
\includegraphics[width=0.7\linewidth]{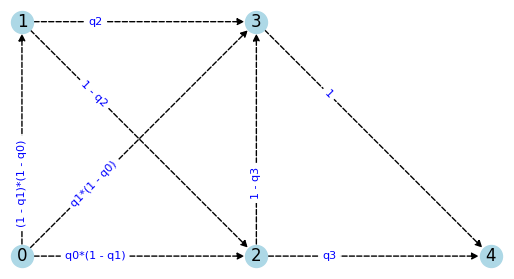}
\caption{Example of path. The path is a directed graph consisting of five nodes from 0 to 4. Some nodes are connected by edges. Among the combinations of these edges, the edge with the combination that minimizes the cost is selected to find a single path from 0 to 4.}
\label{fig:path_example}
\end{figure}

The path to be compressed (e.g., Fig.~\ref{fig:path_example}, below) is a directed graph. Suppose that the nodes in the graph are numbered in order from 0, and that connections are possible from the $i$ th node to up to $k_i$ nodes forward. For example, if the 0 th node can be connected to three possible forward nodes, and a path is chosen that connects the 0 th node to the 2nd node, the 1st node is dropped and the data is compressed by one node. Hereafter, a node is also called a ``vertex'' and a connection is called an ``edge.'' The set consisting of all vertices is denoted as $V$, and the set consisting of all edges is denoted as $E$, which is a subset of $V \times V$.

For simplicity, consider the path in Fig.~\ref{fig:path_example} as an example. In the following, three edges extend forward from the 0 th vertex, two edges extend forward from the first and second vertices, and the only possible connection from the third vertex is to the fourth vertex, which is a single edge. In this case $V = \{0, 1, 2, 3, 4\}$ and $E=\{(0,1), (0,2), (0,3), (1,2), (1,3), (2,3), (2,4)\}$.

The proposed method compresses data by solving a combinatorial optimization such that the sum of the weights of the selected edges is minimized by giving all edges a weight of 1.

\subsection{Formulation in Combinatorial Optimization---QUBO and HOBO}

Solving a combinatorial optimization problem in QAOA requires formulating the problem. Typical formulations include QUBO (Quadratic Unconstrained Binary Optimization) and HOBO (Higher-Order Binary Optimization).

QUBO is a mathematical framework for expressing optimization problems:
\begin{align}
f (x) = \sum_i a_i x_i + \sum_{i < j} b_{i j} x_i x_j
\end{align}
where $x_i \in \{0, 1\}$ and $a_i, b_{i j} \in \R$.

To implement the problem formulated in QUBO on a quantum computer, each variable $x_i$ is replaced by $\frac{1}{2} (1 - Z_i)$, where $Z_i$ is the Pauli-Z operator acting on the $i$ th qubit. In this way, QUBO is represented as a Hamiltonian to change the state of qubits.

HOBO is an extension of QUBO and is represented by a polynomial containing terms of degree three or higher:
\begin{align}
f (x) = \sum_i a_i x_i + \sum_{i < j} b_{i j} x_i x_j + \sum_{i < j < k} c_{i j k} x_i x_j x_k + \cdots
\end{align}

In implementing the problem formulated in HOBO on a quantum computer, each variable $x_i$ is replaced by $\frac{1}{2} (1 - Z_i)$, as in the QUBO case.

\subsection{HOBO vs QUBO} \label{sec:hobo_qubo}

We would like to make a comparison between the formulation with the HOBO and the formulation with the QUBO in terms of the binary variables required.

First, we note the number of binary variables required in the HOBO formula. The number of edges $e_i$ connected from a vertex $i$ to a forward vertex is represented as
\begin{align}
e_i = \#\{j; (i, j) \in E\} = \# E_i
\end{align}
where $E_i$ is the set of edges such that the vertex $i$ is in front. The number of binary variables needed to represent this value is $\lceil \log_2 e_i \rceil$, where $\lceil x \rceil$ is the smallest integer greater than or equal to $x$. Thus, the total number of binary variables required for the formulation in the HOBO formula is expressed as follows
\begin{align} \label{eq:hobo_binary}
\sum_{i \in V} \lceil \log_2 e_i \rceil
\end{align}

Next, we examine the QUBO formula. Define a binary variable $y_{(i,j)}$ such that it takes $1$ when the edge $(i,j) \in E$ is connected and $0$ otherwise. In this case, the QUBO formula becomes Eq.~(\ref{eq:qubo}).

\begin{align} \label{eq:qubo}
\begin{split}
\ &\underbrace{\sum_{(i,j) \in E} y_{(i,j)}}_{\text{Cost function}} \\
+ &\underbrace{\lambda_1 \sum_{j \in V} \left(\sum_{(i,j) \in E} y_{(i,j)} - \sum_{(j,k) \in E} y_{(j,k)}\right)^2}_{\text{Soft constraint \#1}} \\
+ &\underbrace{\lambda_2 \left( \left( \sum_{(0,j) \in E} y_{(0,j)} - 1 \right)^2 + \left( \sum_{(i,N-1) \in E} y_{(i,N-1)} - 1 \right)^2 \right)}_{\text{Soft constraint \#2}}
\end{split}
\end{align}

The cost function minimizes the number of edges to be selected. Soft constraint \#1 requires that all vertices have the same number of incoming and outgoing edges. Soft constraint \#2 requires that only one edge through the start and end vertices be exclusively selected. Thus, the total number of binary variables required for the formulation in the QUBO formula is expressed as follows
\begin{align} \label{eq:qubo_binary}
\# E = \sum_{i=0}^{N-1} \# E_i = \sum_{i \in V} e_i
\end{align}
where $E_i$ is the set of edges such that the vertex $i$ is in front.

Comparing Eq.~(\ref{eq:hobo_binary}) and Eq.~(\ref{eq:qubo_binary}), noting that in $x > 0$ and $\log_2 x < x$, the growth in the number of binary variables is much slower when formulated with the HOBO formula.

In our proposed method, we use HOBO based on the above.

\subsection{HOBO formulation by an example}

We would like to consider a HOBO formulation of the graph in Fig.~\ref{fig:path_example}. First, consider assigning integers to the edges connecting vertices. For the edges starting at the $i$ th vertex, the values are
\begin{itemize}
    \item 0 if $(i, i+1) \in E$
    \item 1 if $(i, i+2) \in E$
    \item 2 if $(i, i+3) \in E$
\end{itemize}

In binary numbers, these are \texttt{00}, \texttt{01}, and \texttt{10}, respectively. Consider expressing this in binary variables.

Specifically, consider the $0$ th vertex. Suppose there are binary variables $x_0$ and $x_1$ associated with this vertex. Assume that $x_0=0$ and $x_1=0$, then the edge has the value \texttt{00} and is connected to the vertex $1$. Therefore, the HOBO formula for this edge is
\begin{align}
w_{0, 1} (1 - x_0) (1 - x_1)
\end{align}
where $w_{0, 1}$ is the weight of the edge connecting vertex $0$ and vertex $1$, which is always 1 this time. Similarly, the HOBO formula for the edge connecting vertex $0$ and vertex $2$ is as follows, assuming that the edge has the value \texttt{01} with $x_0=1$ and $x_1=0$,
\begin{align}
w_{0, 2} x_0 (1 - x_1)
\end{align}

The connection between vertex $0$ and vertex $3$ is represented as follows
\begin{align}
w_{0, 3} (1 - x_0) x_1
\end{align}

By the way, in the present example, the cases $x_0=1$ and $x_1=1$ have no vertices to be connected. Therefore, this case is considered as one of \textit{constraint terms}. In other words, the following is a constraint term:
\begin{align}
x_0 x_1
\end{align}

Next we look at vertex $1$, since the connections from vertex $1$ are to vertices $2$ or $3$, only one binary variable is needed, say $x_2$. Suppose that it is connected to vertex $2$ with $x_2 = 0$ and to vertex $3$ with $x_2 = 1$. Let $w_{1, 2}$ be the weight when vertex $1$ is connected to vertex $2$, then for this to count as a cost, not only vertex $1$ must be connected to vertex $2$, but vertex $0$ and vertex $1$ must also be connected and the path connected. Thus, the HOBO formula including $w_{1, 2}$ is as follows:
\begin{align}
w_{1, 2} (1 - x_0) (1 - x_1)(1 - x_2)
\end{align}

In general, when the HOBO formula $h_{i, j}$ includes the weights $w_{i, j}$, $h_{i, j}$ satisfies the following recurrence formula:
\begin{align}
\begin{split}
h_{i, j} = \left(\sum_{(k, i) \in E} h_{k, i}\right) \cdot \Big[ x_{i, 1}^{b_1} (1 - x_{i, 1})^{1 - b_1} \times \cdots \times x_{i, m}^{b_m} (1 - x_{i, m})^{1 - b_m}\Big]
\end{split}
\end{align}

Let $x_{i, 1},\ldots, x_{i, m}$ be a binary variable used to represent the forward connection from vertex $i$, where edge $(i, j)$ is assumed to be connected with binary $b_1 \cdots b_m$. Also, $E$ is the set of all existing edges.

If we proceed with the HOBO formulation by paying attention to the recurrence formula, the HOBO formula corresponding to the graph in Fig.~\ref{fig:path_example} becomes Eq.~(\ref{eq:hobo}). where $W$ is the weight for the constraint terms, which is assumed to be sufficiently large.
\begin{align} \label{eq:hobo}
\begin{split}
f (x) &= \sum_{(i, j) \in E} w_{i, j} h_{i, j} \\
&= w_{0, 1} (1 - x_0) (1 - x_1) + w_{0, 2} x_0 (1 - x_1) + w_{0, 3} (1 - x_0) x_1 + w_{1, 2} (1 - x_0) (1 - x_1)(1 - x_2) \\
&\ + w_{1, 3} (1 - x_0) (1 - x_1) x_2 + w_{2, 3} \{x_0 (1 - x_1) + (1 - x_0) (1 - x_1)(1 - x_2)\} (1 - x_3) \\
&\ + w_{2, 4} \{x_0 (1 - x_1) + (1 - x_0) (1 - x_1)(1 - x_2)\} x_3 \\
&\ + w_{3, 4} [(1 - x_0) x_1 + (1 - x_0) (1 - x_1) x_2 + \{x_0 (1 - x_1) + (1 - x_0) (1 - x_1)(1 - x_2)\} (1 - x_3)] \\
&\ + W (x_0 x_1)
\end{split}
\end{align}

\section{Experiment} \label{sec:experiment}

\subsection{Edge thinning \& deploy} \label{sec:edge_thinning}

A possible preprocessing step is to discard edges with large deviations from the path when connecting edges to the forward vertices.

\begin{figure}[!ht]
\centering
\includegraphics[width=0.8\linewidth]{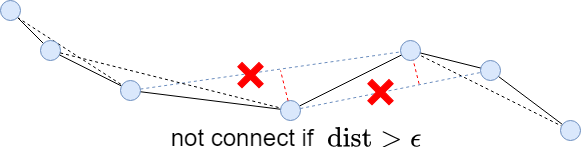}
\caption{Example of preprocess. Connect edges from each node to forward nodes. However, if there is an intermediate node between endpoints whose distance to the edge exceeds a certain threshold $\epsilon$, the edge is discarded.}
\label{fig:edge_thinning}
\end{figure}

This allows the removal of edges that deviate significantly from the shape of the path and reduces the computational cost of combinatorial optimization.

In the proposed method, this edge thinning process is performed as a preprocessing step when applied to actual route data.

Note that this is not a global process like Sec.~\ref{sec:overview_of_rdp}, but a local process along the route.

\subsection{Proposed method}

After the aforementioned edge thinning, the proposed method compresses the data by solving a combinatorial optimization problem using the HOBO formulation. The pseudo code is shown in Appendix~\ref{sec:rdp_pseudo_code}, Alg.~\ref{alg:our_method}.

\subsection{Experimental Result}

All experiments were performed using the following Qiskit SDK.

\begin{itemize}
    \item qiskit 1.2.0
    \item qiskit-aer 0.15.0
    \item qiskit\_ibm\_runtime 0.28.0
\end{itemize}

\begin{figure}[!ht]
\centering
\includegraphics[width=0.85\linewidth]{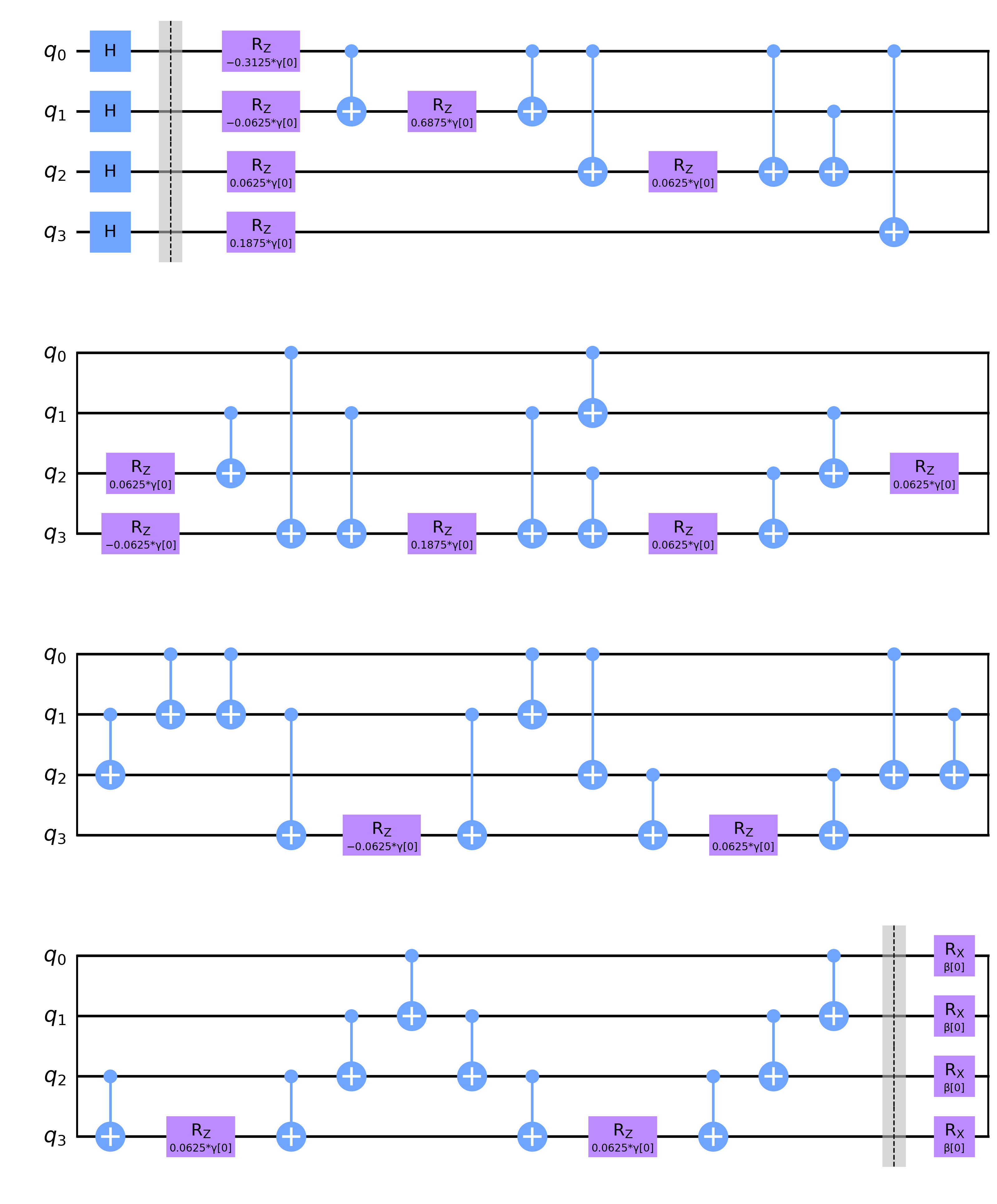}
\caption{The quantum circuit for the toy problem}
\label{fig:circuit}
\end{figure}

First, we show the experimental results on the actual quantum computer \texttt{ibm\_brisbane} for the route data in Fig.~\ref{fig:path_example}. This problem is a toy problem position, and the edge thinning of Sec.~\ref{sec:edge_thinning} is \textit{not} applied. The implementation of Fig.~\ref{fig:path_example} in a quantum circuit is shown in Fig.~\ref{fig:circuit}. In this toy problem, only the Pauli-Z rotation gate \texttt{RZ} and the $Z \otimes Z$ rotation gate \texttt{RZZ} appear, but in the optimization using more complex actual route data, rotated versions of the gates corresponding to more Kronecker products in \texttt{Z} are used. Regarding the implementation of such rotation gates, useful references include the method based on parity computation with an ancilla qubit to determine the phase rotation (see Fig. 4.19 in \cite{Nielsen_Chuang_2010}), as well as ancilla-qubit-free approaches described in \cite{Seeley_Richard_Love_2012,Glos_Krawiec_Zimboras_2022,Verchere2023}. Although both methods can be implemented using a combination of a sequence of staircase-like \texttt{CX} gates and a single \texttt{RZ} gate, in this work we employed the ancilla-qubit-free implementation of \cite{Seeley_Richard_Love_2012}.

\begin{figure}[!ht]
\centering
\includegraphics[width=0.7\linewidth]{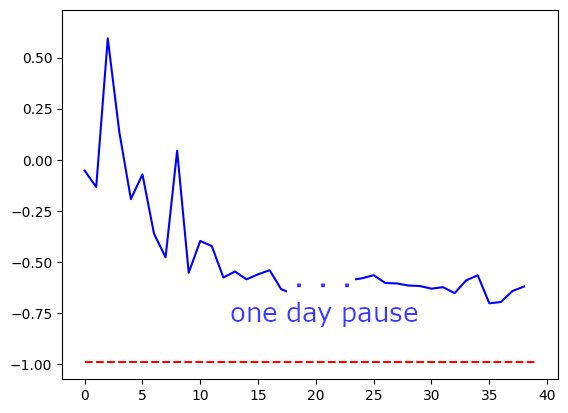}
\caption{Transitions in Hamiltonian expectation values. A timeout occurred while connecting to \texttt{ibm\_brisbane} during execution. The process was resumed from the intermediate results the following day to obtain the final results.}
\label{fig:real_device_cobyla}
\end{figure}

\begin{figure}[!ht]
\centering
\includegraphics[width=0.7\linewidth]{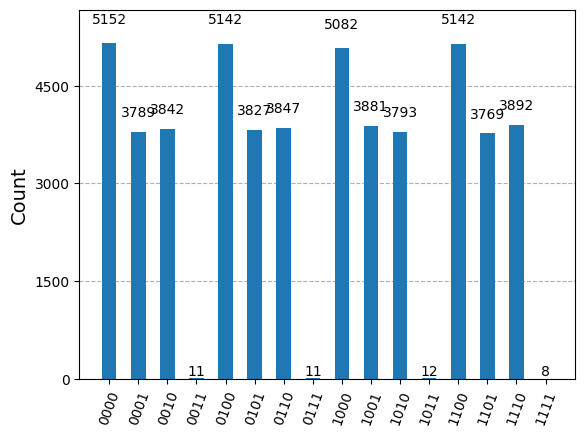}
\caption{Sampling results in Aer simulator. The case where node 0 is not connected to any other node, i.e., $q_0 = 1$ and $q_1 = 1$, has been rarely sampled.}
\label{fig:sampling_result}
\end{figure}

\begin{figure}[!ht]
\centering
\includegraphics[width=0.7\linewidth]{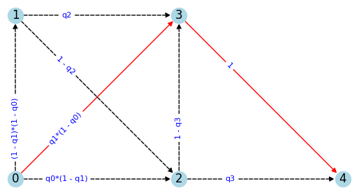}
\caption{Compressed paths corresponding to sampling results}
\label{fig:selected_path}
\end{figure}

To make training more efficient, the initial values for \texttt{RZ} and \texttt{RZZ} are not random parameters, but values that were previously optimized for 11 iterations of SLSQP on the simulator. Prior to optimization on the actual quantum computer, as a preliminary experiment, we loaded the information of \texttt{ibm\_brisbane} into the Aer simulator and tested the optimization methods COBYLA, Powell, SLSQP, and SPSA. Since the results of COBYLA were good, we decided to optimize with COBYLA and apply error mitigation methods to the actual machine. T-REX (Twirled Readout Error eXtinction) and Pauli-twirling were used as error mitigation methods. The parameters obtained as a result of the optimization were sampled with the Aer simulator. The optimization process and the sampling results are shown in Fig.~\ref{fig:real_device_cobyla} and Fig.~\ref{fig:sampling_result}, respectively. Among the quantum states obtained, selecting the one that reduces the data size the most resulted in the data compression as shown in Fig.~\ref{fig:selected_path}.

\begin{figure}[!ht]
\centering
\includegraphics[width=0.8\linewidth]{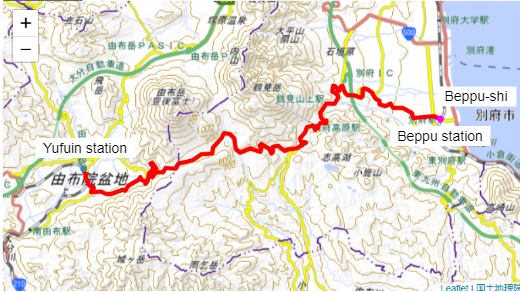}
\caption{Route from Beppu Station to Yufuin Station drawn with GPXEdit~\cite{GPXEdit_2021} from the 2024 map data (Geographical Survey Institute map (Denshi Kokudo Web~\cite{Kokudo_Chiriin_2024})).}
\label{fig:yufuin_path}
\end{figure}

Next, we will discuss optimization using actual route data. In this case, we consider the route (red line) shown in Fig.~\ref{fig:yufuin_path} from Beppu Station to Yufuin Station in Oita Prefecture.

The route coordinate data was obtained by Google Cloud's Directions API.

Since this route is long and difficult to optimize on a NISQ quantum computer, a simulator is used to perform the calculations. In addition, as will be explained in detail in Sec.~\ref{sec:discussion}, the depth of the quantum circuit may increase exponentially if the path is long. For this reason, we will consider dividing the pathway accordingly.

\begin{figure}[!ht]
\centering
\includegraphics[width=0.7\linewidth]{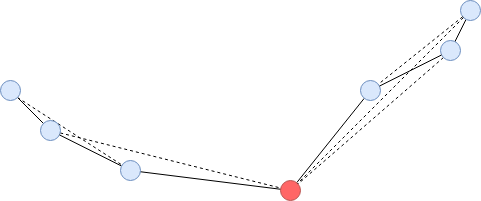}
\caption{Theoretical division point. The path bends sharply at the red node. Any edge that includes such nodes as intermediate nodes is discarded during preprocessing. Therefore, this red node is always marked and retained in the final result, and it is possible to divide the path at this point and perform optimization separately.}
\label{fig:division_point}
\end{figure}

At bends in the path, edges with large deviations from the path are thinned out by preprocessing as described in Sec.~\ref{sec:edge_thinning}. Therefore, a node may appear where there is no edge that straddles it, such as the red one in the figure. Since this node is always passed and selected, the graph can be cut at this node and the results of the two separated optimization problems can be combined later. Such a node is specifically called a \textit{theoretical division point}.

In addition to theoretical division points a path may be cut off as necessary for computational convenience. For example, a path may exceed a certain length and become difficult to calculate in the simulator since the depth of the corresponding circuit is too deep. Nodes at which a path is divided for such a reason are called \textit{computational division points}.

\begin{figure*}[!ht]
\centering
\includegraphics[width=\linewidth]{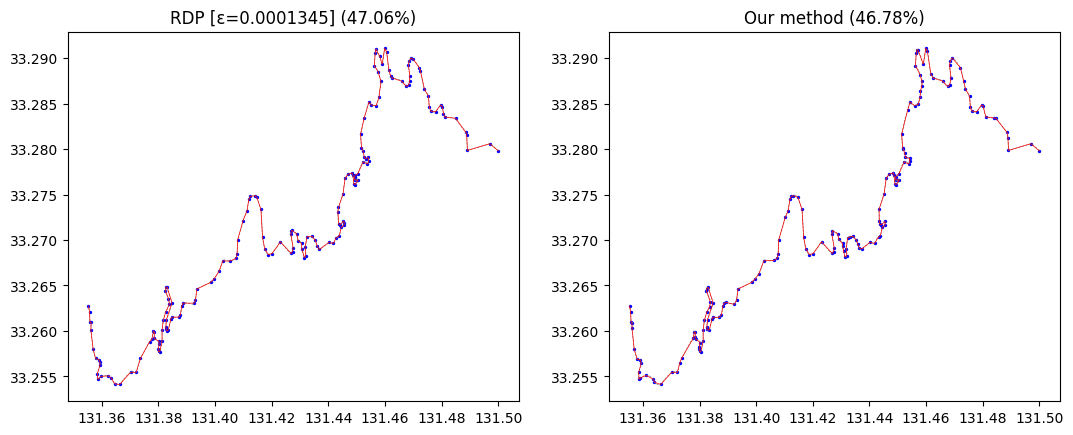}
\caption{Comparison of RDP and our method for Yufuin route. The proposed method achieves a slightly better compression result.}
\label{fig:rdp_vs_proposed_yufuin}
\end{figure*}

Fig.~\ref{fig:rdp_vs_proposed_yufuin} presents the results for edge thinning of the RDP and the proposed method when $\epsilon = 0.0001345$ is chosen. Here, the choice of the value $\epsilon = 0.0001345$ has no particular significance.

In this experiment, the size after compression is 47.06\% of the original size for RDP and 46.78\% for the proposed method, indicating that the proposed method has a slightly higher compression ratio.

The details of the nodes are also shown in Table~\ref{table:yufuin_node_detail}. As can be seen from the table, the computational division points account for 4.76\% of the total nodes, around which the selection of the optimal path may be hampered.

\begin{table}[!ht]
\centering
\caption{Selected points are divided into three categories: Normal, Theoretical and Computational division points. The percentage is the ratio to the total points.}
\begin{tabular}{ccc}
\toprule
& \textbf{RDP} & \textbf{Proposed} \\
\midrule
Total points & 357 & 357 \\
\midrule
Selected points & 168 (47.06\%) & \textbf{167} (46.78\%) \\
\midrule
\multicolumn{1}{r}{Normal} & - & 84 (23.53\%) \\
\multicolumn{1}{r}{Theor. Div.} & - & 66 (18.49\%) \\
\multicolumn{1}{r}{Comp. Div.} & - & 17 (4.76\%) \\
\midrule
Dropped points & 189 (52.94\%) & \textbf{190} (53.22\%) \\
\bottomrule
\end{tabular}
\label{table:yufuin_node_detail}
\end{table}

\begin{figure}[!ht]
\centering
\includegraphics[width=0.8\linewidth]{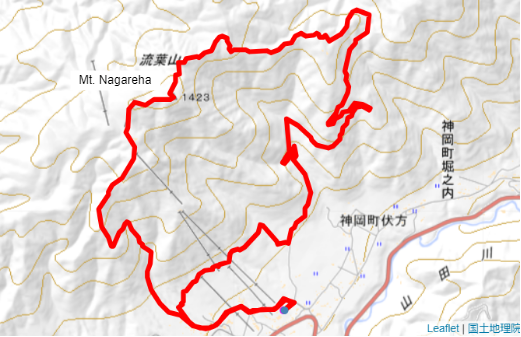}
\caption{Route of The 11th Okuhida Trail Run Short 12km Course drawn with GPXEdit~\cite{GPXEdit_2021} from the 2024 map data (Geographical Survey Institute map (Denshi Kokudo Web~\cite{Kokudo_Chiriin_2024})).}
\label{fig:okuhida_path}
\end{figure}

We would like to discuss optimization using route data other than the above. Here, we consider the route (red line) shown in Fig.~\ref{fig:okuhida_path}, which is the course of a trail running event held in Hida City, Gifu Prefecture.

The route data were those published in \cite{Okuhida_Trail_Run_11th_2024}.

Because the route is also long and difficult to optimize with the NISQ quantum computer, so again, a simulator is used.

\begin{figure*}[!ht]
\centering
\includegraphics[width=\linewidth]{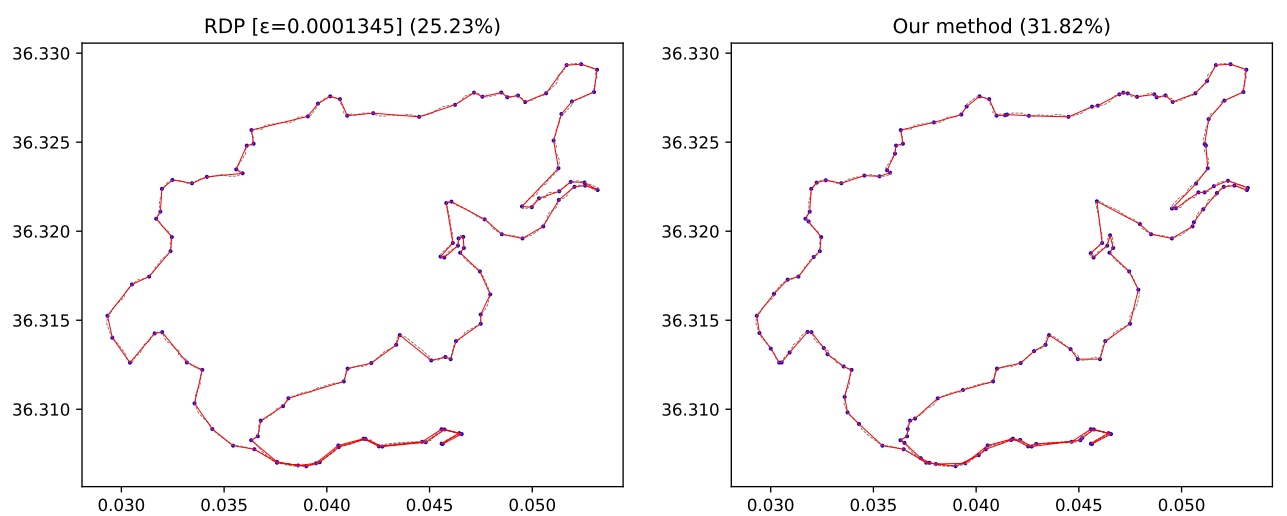}
\caption{Comparison of RDP and our method for Okuhida route. In this case, RDP achieves a better compression result.}
\label{fig:rdp_vs_proposed_okuhida}
\end{figure*}

Fig.~\ref{fig:rdp_vs_proposed_okuhida} presents the results for the RDP and the proposed method when $\epsilon = 0.0001345$ is chosen as in the previous case.

The details of the nodes are shown in Table~\ref{table:okuhida_node_detail}. As can be seen from the table, the computational division points account for 13.18\% of the total nodes, around which the optimal path selection may be hampered.

As a result, the RDP compression ratio is better in this case. One reason is that the proposed method has 13.18\% of computational division points compared to 4.76\% for the previous Yufuin route, and 58 nodes may be preventing proper path selection. The paths were relatively gently circumscribed, with few sharp bends, and thus theoretical division points were unlikely to have occurred. This resulted in relatively long partial paths and computationally difficult graphs, which required a large number of computational division points to make them computable.

\begin{table}[!ht]
\centering
\caption{Selected points are divided into three categories: Normal, Theoretical and Computational division points. The percentage is the ratio to the total points.}
\begin{tabular}{ccc}
\toprule
& \textbf{RDP} & \textbf{Proposed} \\
\midrule
Total points & 440 & 440 \\
\midrule
Selected points & \textbf{111} (25.23\%) & 140 (31.82\%) \\
\midrule
\multicolumn{1}{r}{Normal} & - & 61 (13.86\%) \\
\multicolumn{1}{r}{Theor. Div.} & - & 21 (4.77\%) \\
\multicolumn{1}{r}{Comp. Div.} & - & 58 (13.18\%) \\
\midrule
Dropped points & \textbf{329} (74.77\%) & 300 (68.18\%) \\
\bottomrule
\end{tabular}
\label{table:okuhida_node_detail}
\end{table}

\begin{figure}[!ht]
\centering
\includegraphics[width=0.8\linewidth]{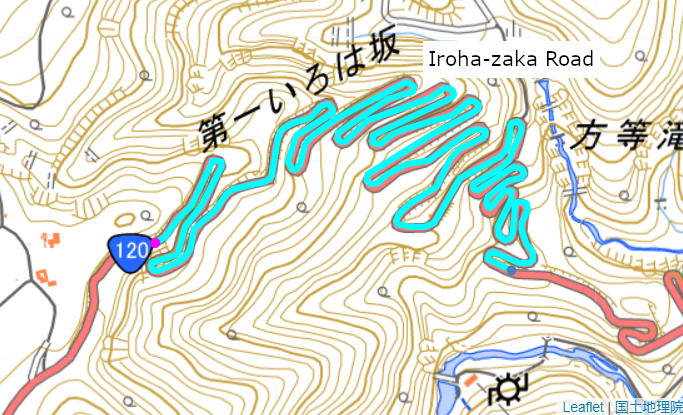}
\caption{Route of the Iroha-zaka Road drawn with GPXEdit~\cite{GPXEdit_2021} from the 2024 map data (Geographical Survey Institute map (Denshi Kokudo Web~\cite{Kokudo_Chiriin_2024})).}
\label{fig:iroha_path}
\end{figure}

Finally, we would like to look at the third case. This time, we deal with the ``Iroha zaka Road'' route (blue) in Tochigi Prefecture, as shown in Fig.~\ref{fig:iroha_path}. The coordinate data of the path was again obtained using Google Cloud's Directions API.

\begin{figure*}[!ht]
\centering
\includegraphics[width=\linewidth]{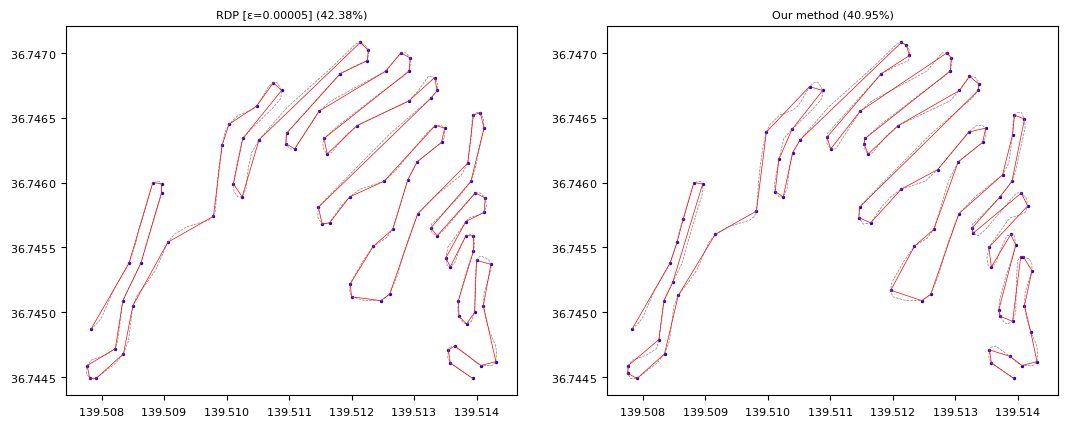}
\caption{Comparison of RDP and our method for Iroha-zaka Road route. The proposed method achieves a slightly better compression result.}
\label{fig:rdp_vs_proposed_iroha}
\end{figure*}

Fig.~\ref{fig:rdp_vs_proposed_iroha} presents the results for the RDP and the proposed method when $\epsilon = 0.00005$ is chosen. Note that this is a different value from the previous two experiments.

In this experiment, the size after compression is 42.38\% of the original size for RDP and 40.95\% for the proposed method, indicating that the proposed method has a slightly higher compression ratio.

The details of the nodes are also shown in Table~\ref{table:iroha_node_detail}. As can be seen from the table, the computational division points account for 0.95\% of the total nodes, around which the selection of the optimal path may be hampered.

The values of $\epsilon$ in the three experiments so far ware set arbitrarily. Here, we would like to give an overview of the relationship between the various $\epsilon$ values and the compression ratio in Fig.~\ref{fig:proposed_to_RDP} in Sec.~\ref{sec:discussion}.

\begin{table}[!ht]
\centering
\caption{Selected points are divided into three categories: Normal, Theoretical and Computational division points. The percentage is the ratio to the total points.}
\begin{tabular}{ccc}
\toprule
& \textbf{RDP} & \textbf{Proposed} \\
\midrule
Total points & 210 & 210 \\
\midrule
Selected points & 89 (42.38\%) & \textbf{86} (40.95\%) \\
\midrule
\multicolumn{1}{r}{Normal} & - & 58 (27.62\%) \\
\multicolumn{1}{r}{Theor. Div.} & - & 26 (12.38\%) \\
\multicolumn{1}{r}{Comp. Div.} & - & 2 (0.95\%) \\
\midrule
Dropped points & 121 (57.62\%) & \textbf{124} (59.05\%) \\
\bottomrule
\end{tabular}
\label{table:iroha_node_detail}
\end{table}

\section{Discussion} \label{sec:discussion}

\noindent\textbf{Path combinations:} We have discussed how to select the path with the fewest number of nodes from path candidates. Here we consider the total number of path combinations. For simplicity, let us assume that there is a path $\{0, 1, \ldots, n-1\}$ with $n \geq 3$ vertices. We would like to consider a combination of paths that reach from the start point to the end point by skipping at most one node. For example, in the case of $n=5$, the path is $\{0, 1, 2, 3, 4\}$, and if one skipping is allowed, $\{0, 2, 3, 4\}$ etc. can be considered. If skipping is allowed multiple times, $\{0, 2, 4\}$ is also valid. Of course, $\{0, 1, 2, 3, 4\}$, which does not skip at all, is also a combination.

Let $C_i$ be the number of combinations that reach the vertex $i$ from the starting vertex $0$. Then, we have the following recurrence relation:
\begin{align}
C_i = C_{i-1} + C_{i-2}
\end{align}
This is because the combination reaching $i$ is the sum of the combination reaching $i-1$ plus the path from $i-1$ to $i$ and the combination reaching $i-2$ plus the path from $i-2$ to $i$. Since $C_0=1$, $C_1=1$, $\{C_i\}_{0 \leq i \leq n - 1}$ is the so-called Fibonacci sequence. Therefore, the general term is
\begin{align} \label{eq:approx_fibonacci}
C_i = \frac{1}{\sqrt{5}} \left( \left(\frac{1+\sqrt{5}}{2}\right)^{i+1} - \left(\frac{1-\sqrt{5}}{2}\right)^{i+1} \right),
\end{align}
($0 \leq i \leq n-1$). Thus, when $n$ is sufficiently large and $i$ is sufficiently large, we have approximately
\begin{align}
\begin{split}
C_i &= \frac{1}{\sqrt{5}} \left( \left(\frac{1+\sqrt{5}}{2}\right)^{i+1} \left(1 - \left(\frac{1-\sqrt{5}}{1+\sqrt{5}}\right)^{i+1} \right) \right) \\
&\approx \frac{1}{\sqrt{5}} \left(\frac{1+\sqrt{5}}{2}\right)^{i+1}.
\end{split}
\end{align}

Based on the above estimation, we consider the proposed method. Consider a path with $n$ vertices $\{0, 1, \ldots, n-1\}$, whose nodes can be connected to at most two forward vertices. In this case, the combination reaching the $i$ th vertex is $C_i$ and is approximately given by Eq.~(\ref{eq:approx_fibonacci}). The highest order term of the Ising equation for each path yields one gate of the form $R Z_0 \otimes Z^{\otimes k} \otimes Z_{n-2}$ where $k \leq n - 1$. Since these are exclusively arranged, the depth of the corresponding quantum circuit is at least of the order of $\frac{1}{\sqrt{5}} \left(\frac{1+\sqrt{5}}{2}\right)^{n}$. Hence, it is suggested that when $n$ is sufficiently large, this quantum circuit will be exponentially deep. Similarly, in other cases, it is generally expected that the longer the path to be optimized, the exponentially deeper the corresponding quantum circuit will be.

\noindent\textbf{Possible advantage:} As we have seen above, there is a possibility that the proposed method can compress better than RDP depending on the path. One possible reason why the proposed method may have a better compression ratio than RDP is that RDP only performs global optimization using $\epsilon$. On the other hand, the proposed method can perform global optimization by preprocessing edge thinning and local optimization by combinatorial optimization at the same time.

\noindent\textbf{Limitations:} In the Okuhida example, the non-essential path dividision (i.e., at computational division points) resulted in poorer compression than RDP. As seen in the above example, the current situation is such that large-scale optimization is difficult due to the constraints of computational resources and the need to select computationally appropriate division points. However, with the development of quantum computers in the future, it is expected that better results will become possible to perform calculations with a larger number of qubits and to discover larger-scale optimization algorithms.

\noindent\textbf{Other applications:} The proposed method is also flexible. Although not addressed in this study, for example, the HOBO formulation can be adjusted according to the characteristics of the route and/or requirements of the optimization. For example, if data compression is not the main concern, but rather the shortest possible path distance, the edge weights can be taken as the distance between the vertices. In this way, the system can be adapted to various application scenarios and is considered to be highly practical.

\begin{figure*}[!ht]
\centering
\includegraphics[width=\linewidth]{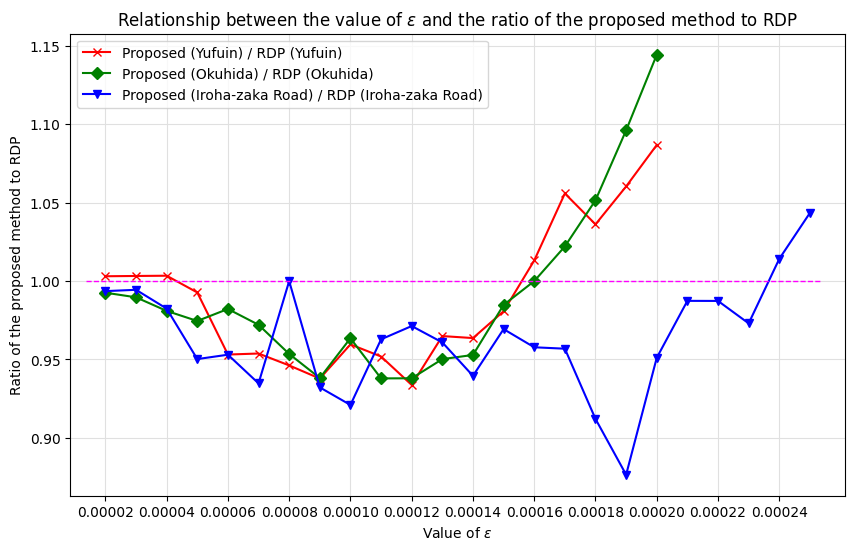}
\caption{Relationship between the value of $\epsilon$ and the ratio of the proposed method to RDP (or the number of points after compression by the proposed method divided by the number of points after compression by RDP). To match the scale, each route dataset is scaled so that the average distance between nodes becomes 0.000653. In all cases, both methods perform at the same level when $\epsilon$ is small because neither can achieve significant compression. As $\epsilon$ gradually increases, the proposed method gains an advantage. However, beyond approximately $\epsilon > 0.00016$, RDP becomes superior. This is likely due to computational resource limitations, which prevent the proposed method from executing large-scale optimizations, leading to an increase in computational division points and a decrease in compression efficiency.}
\label{fig:proposed_to_RDP}
\end{figure*}

\textbf{Relationship between the value of $\epsilon$ and the compression ratio:} We would like to confirm the relationship between the value of $\epsilon$ and the compression ratio for the Yufuin, Okuhida Trail Run Course, and Irohazaka Road routes, and compare the results of RDP and the ideal solution obtained by brute-force calculation of the optimization part of the proposed method on a classical computer. First, since the distance between points in the route data varies from route to route, we aligned the data with the Yufuin data to match the scale\footnote{The average distances between points for the Yufuin, Okuhida, and Irohazaka Road data were 0.000653, 0.000291, and 0.000173, respectively, so these were all scaled to an average of 0.000653}. Using the scaled route data, the proposed method and RDP were run for each $\epsilon$, and the ratio of the number of points after compression by the proposed method to that by RDP is shown in Fig.~\ref{fig:proposed_to_RDP}.

\begin{figure}[!ht]
\centering
\includegraphics[width=0.5\linewidth]{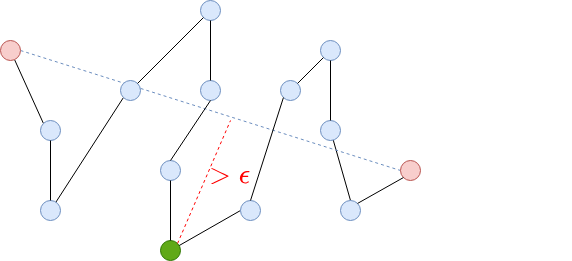}
\caption{RDP's first step for a zigzag road. The furthest point C (green) from the line segment with the first point A (red) and the last point B (red) is marked and retained.}
\label{fig:possible_rdp_weakness-1}
\end{figure}

\begin{figure}[!ht]
\centering
\includegraphics[width=0.5\linewidth]{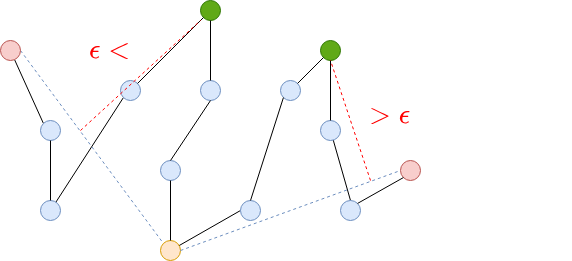}
\caption{RDP's subsequent steps for a zigzag road. Even as the recursive process progresses, it is relatively easy to find subpaths that are nearly orthogonal to the ``line segment.'' Therefore, in a zigzag road, points that are at least $\epsilon$ away from the ``line segment'' are more likely to be found, and it is expected that these points will tend to be marked and retained.}
\label{fig:possible_rdp_weakness-2}
\end{figure}

When $\epsilon$ is small, few nodes can be thinned out, and both RDP and the proposed method can hardly compress them. When $\epsilon$ is increased to some extent, the proposed method has an advantage. This may be due to the contribution of local optimization described in ``Possible advantage''. When $\epsilon$ is sufficiently large, RDP becomes dominant. As described in ``Limitations'', the proposed method is difficult to perform largescale optimization due to the limitation of computational resources. Therefore, as $\epsilon$ is increased, the non-essential path division increases and the compression efficiency decreases. Fig.~\ref{fig:proposed_to_RDP} seems to suggest that the proposed method may be superior for some routes. For example, the Irohazaka data shows better compression than RDP for more $\epsilon$. One possibility is that the path is zigzagging. In the case of zigzag paths, we find many parts that are nearly orthogonal to the ``line segments'' in the RDP algorithm. Thus, if $\epsilon$ is less than the distance between adjacent points, one would imagine that more points are more likely to be marked and retained (Fig.~\ref{fig:possible_rdp_weakness-1}, \ref{fig:possible_rdp_weakness-2}). For this reason, compared to relatively simple routes such as Yufuin and Okuhida, zigzagging routes such as the Irohazaka may allow the proposed method to compete with RDP even with a relatively large $\epsilon$ and with restrictions.

\section{Conclusion} \label{sec:conclusion}

In this study, we propose a quantum optimization-based route compression technique. The proposed method compresses data by finding the optimal combination of paths using QAOA (Quantum Approximate Optimization Algorithm). We formulate the problem using Higher-Order Binary Optimization (HOBO) to treat it as a higher-dimensional optimization problem and achieve efficient compression while reducing the number of binary variables.

Experimental results show that the proposed method can compress slightly better than RDP depending on the path and the $\epsilon$ value. Other than the compression ratio, the proposed method has the advantage of being flexible according to the shape and characteristics of the route. For example, by adjusting the HOBO formulation according to the characteristics of the route and/or requirements of the optimization, the proposed method can be used not only for data compression, but also for other purposes such as minimizing the distance of the route.

On the other hand, the current limitations of computational resources make large-scale optimization difficult. Future development of quantum computers will make it possible to perform calculations with a larger number of qubits, and the discovery of larger-scale optimization algorithms is expected to lead to better results.

We intend to further develop the proposed method and apply it to practical applications. For example, we plan to verify the effectiveness of the proposed method in areas where combinatorial optimization plays an important role, such as car navigation route finding and delivery planning problems. As the performance of quantum computers improves, we would also like to develop an improved compression method for larger routes.

\section*{Acknowledgments}
Takao Tomono is supported by the Center of Innovations for Sustainable Quantum AI (JST), Grant No. JPMJPF2221.

\bibliographystyle{plain}
\bibliography{main}

\begin{thebibliography}{10}

\bibitem{GPXEdit_2021}
Gpxedit (japanese), 2021.
\newblock A web-based GPX editor.

\bibitem{Kokudo_Chiriin_2024}
Kokudo chiriin (geospatial information authority of japan), 2024.
\newblock The national mapping agency of Japan.

\bibitem{Okuhida_Trail_Run_11th_2024}
Okuhida trail run (11th) (japanese), 2024.
\newblock A trail run event in Japan. The 11th Okuhida Trail Run page is in \url{https://www.actrep-sports.com/}.

\bibitem{Basile2017}
Ivano Basile and Fabio Tamburini.
\newblock Towards quantum language models.
\newblock In {\em Proceedings of the 2017 Conference on Empirical Methods in Natural Language Processing}, pages 1840--1849. Association for Computational Linguistics, September 2017.

\bibitem{Bausch2020}
Johannes Bausch.
\newblock Recurrent quantum neural networks.
\newblock In H.~Larochelle, M.~Ranzato, R.~Hadsell, M.F. Balcan, and H.~Lin, editors, {\em Advances in Neural Information Processing Systems}, volume~33, pages 1368--1379. Curran Associates, Inc., 2020.

\bibitem{B_rtschi_2019}
Andreas Bärtschi and Stephan Eidenbenz.
\newblock {\em Deterministic Preparation of Dicke States}, pages 126--139.
\newblock Springer International Publishing, 2019.

\bibitem{Cong2019}
Iris Cong, Soonwon Choi, and Mikhail~D. Lukin.
\newblock Quantum convolutional neural networks.
\newblock {\em Nature Physics}, 15(12):1273--1278, 2019.

\bibitem{Douglas_Peucker_1973}
Thomas K.~Peucker David H.~Douglas.
\newblock Algorithms for the reduction of the number of points required to represent a digitized line or its caricature.
\newblock {\em The Canadian Cartographer}, 10(2):112--122, 1973.

\bibitem{Evered2023}
Simon~J. Evered, Dolev Bluvstein, Marcin Kalinowski, Sepehr Ebadi, Tom Manovitz, Hengyun Zhou, Sophie~H. Li, Alexandra~A. Geim, Tout~T. Wang, Nishad Maskara, Harry Levine, Giulia Semeghini, Markus Greiner, Vladan Vuleti\'{c}, and Mikhail~D. Lukin.
\newblock High-fidelity parallel entangling gates on a neutral-atom quantum computer.
\newblock {\em Nature}, 622(7982):268--272, October 2023.

\bibitem{Farhi_Goldstone_Gutmann_2014}
Edward Farhi, Jeffrey Goldstone, and Sam Gutmann.
\newblock A quantum approximate optimization algorithm, 2014.

\bibitem{Fitriya_Purboyo_Prasasti2017}
Luluk~Anjar Fitriya, Tito~Waluyo Purboyo, and Anggumeka~Luhur Prasasti.
\newblock A review of data compression techniques.
\newblock {\em International Journal of Applied Engineering Research}, 12:8956--8963, 01 2017.

\bibitem{Glos_Krawiec_Zimboras_2022}
Adam Glos, Aleksandra Krawiec, and Zoltán Zimborás.
\newblock Space-efficient binary optimization for variational quantum computing.
\newblock {\em npj Quantum Information}, 8:39, 2022.

\bibitem{Goodfellow2014}
Ian Goodfellow, Jean Pouget-Abadie, Mehdi Mirza, Bing Xu, David Warde-Farley, Sherjil Ozair, Aaron Courville, and Yoshua Bengio.
\newblock Generative adversarial nets.
\newblock In {\em Advances in Neural Information Processing Systems}, volume~27. Curran Associates, Inc., 2014.

\bibitem{Hadfield_2019}
Stuart Hadfield, Zhihui Wang, Bryan O’Gorman, Eleanor~G. Rieffel, Davide Venturelli, and Rupak Biswas.
\newblock From the quantum approximate optimization algorithm to a quantum alternating operator ansatz.
\newblock {\em Algorithms}, 12(2):34, February 2019.

\bibitem{Havlicek2019}
Vojtěch Havlíček, Antonio~D. Córcoles, Kristan Temme, Aram~W. Harrow, Abhinav Kandala, Jerry~M. Chow, and Jay~M. Gambetta.
\newblock Supervised learning with quantum-enhanced feature spaces.
\newblock {\em Nature}, 567(7747):209--212, 2019.

\bibitem{He_2016_CVPR}
Kaiming He, Xiangyu Zhang, Shaoqing Ren, and Jian Sun.
\newblock Deep residual learning for image recognition.
\newblock In {\em 2016 IEEE Conference on Computer Vision and Pattern Recognition (CVPR)}, pages 770--778, 2016.

\bibitem{Hershberger_Snoeyink_1992}
John Hershberger and Jack Snoeyink.
\newblock Speeding up the douglas-peucker line-simplification algorithm.
\newblock Technical Report TR-92-07, Department of Computer Science, University of British Columbia, CAN, April 1992.

\bibitem{JAYASANKAR2021}
Uthayakumar Jayasankar, Vengattaraman Thirumal, and Dhavachelvan Ponnurangam.
\newblock A survey on data compression techniques: From the perspective of data quality, coding schemes, data type and applications.
\newblock {\em Journal of King Saud University - Computer and Information Sciences}, 33(2):119--140, 2021.

\bibitem{karras2018progressive}
Tero Karras, Timo Aila, Samuli Laine, and Jaakko Lehtinen.
\newblock Progressive growing of gans for improved quality, stability, and variation.
\newblock In {\em International Conference on Learning Representations}, 2018.

\bibitem{rdp_js}
Marius Karthaus.
\newblock Javascript implementation of the ramer douglas peucker algorithm, 2012.

\bibitem{Kato1950}
Tosio Kato.
\newblock On the adiabatic theorem of quantum mechanics.
\newblock {\em Journal of the Physical Society of Japan}, 5(6):435--439, 1950.

\bibitem{Kim2023}
Youngseok Kim, Andrew Eddins, Sajant Anand, Ken~Xuan Wei, Ewout van~den Berg, Sami Rosenblatt, Hasan Nayfeh, Yantao Wu, Michael Zaletel, Kristan Temme, and Abhinav Kandala.
\newblock Evidence for the utility of quantum computing before fault tolerance.
\newblock {\em Nature}, 618(7965):500--505, June 2023.

\bibitem{Kingma2014}
Diederik~P. Kingma and Max Welling.
\newblock Auto-encoding variational bayes.
\newblock In {\em International Conference on Learning Representations (ICLR) 2014}, 2014.

\bibitem{Krizhevsky_Sutskever_Hinton2012}
Alex Krizhevsky, Ilya Sutskever, and Geoffrey~E Hinton.
\newblock Imagenet classification with deep convolutional neural networks.
\newblock In F.~Pereira, C.J. Burges, L.~Bottou, and K.Q. Weinberger, editors, {\em Advances in Neural Information Processing Systems}, volume~25. Curran Associates, Inc., 2012.

\bibitem{Bo_et_al_2020}
Bo~Liu, Xuechao Liu, Dajun Li, Yu~Shi, Gabriela Fernandez, and Yandong Wang.
\newblock A vector line simplification algorithm based on the douglas–peucker algorithm, monotonic chains and dichotomy.
\newblock {\em ISPRS International Journal of Geo-Information}, 9(4), 2020.

\bibitem{Matsuo2023}
Atsushi Matsuo, Yudai Suzuki, Ikko Hamamura, and Shigeru Yamashita.
\newblock Enhancing vqe convergence for optimization problems with problem-specific parameterized quantum circuits.
\newblock {\em IEICE Transactions on Information and Systems}, E106.D(11):1772--1782, 2023.

\bibitem{Neyens2024}
Samuel Neyens, Otto~K. Zietz, Thomas~F. Watson, Florian Luthi, Aditi Nethwewala, Hubert~C. George, Eric Henry, Mohammad Islam, Andrew~J. Wagner, Felix Borjans, Elliot~J. Connors, J.~Corrigan, Matthew~J. Curry, Daniel Keith, Roza Kotlyar, Lester~F. Lampert, Mateusz~T. M\k{a}dzik, Kent Millard, Fahd~A. Mohiyaddin, Stefano Pellerano, Ravi Pillarisetty, Mick Ramsey, Rostyslav Savytskyy, Simon Schaal, Guoji Zheng, Joshua Ziegler, Nathaniel~C. Bishop, Stephanie Bojarski, Jeanette Roberts, and James~S. Clarke.
\newblock Probing single electrons across 300-mm spin qubit wafers.
\newblock {\em Nature}, 629(8010):80--85, May 2024.

\bibitem{Nielsen_Chuang_2010}
Michael~A. Nielsen and Isaac~L. Chuang.
\newblock {\em Quantum Computation and Quantum Information: 10th Anniversary Edition}.
\newblock Cambridge University Press, 2010.

\bibitem{Peruzzo2014}
Alberto Peruzzo, Jarrod McClean, Peter Shadbolt, Man-Hong Yung, Xiao-Qi Zhou, Peter~J. Love, Alán Aspuru-Guzik, and Jeremy~L. O’Brien.
\newblock A variational eigenvalue solver on a photonic quantum processor.
\newblock {\em Nature Communications}, 5(1):4213, 2014.

\bibitem{RAMER1972244}
Urs Ramer.
\newblock An iterative procedure for the polygonal approximation of plane curves.
\newblock {\em Computer Graphics and Image Processing}, 1(3):244--256, 1972.

\bibitem{Seeley_Richard_Love_2012}
Jacob~T. Seeley, Martin~J. Richard, and Peter~J. Love.
\newblock The bravyi-kitaev transformation for quantum computation of electronic structure.
\newblock {\em The Journal of Chemical Physics}, 137(22):224109, 12 2012.

\bibitem{Simonyan2015}
Karen Simonyan and Andrew Zisserman.
\newblock Very deep convolutional networks for large-scale image recognition.
\newblock In {\em 3rd International Conference on Learning Representations (ICLR 2015)}, pages 1--14, 2015.

\bibitem{Sok20}
Igor~O. Sokolov, Panagiotis~Kl. Barkoutsos, Pauline~J. Ollitrault, Donny Greenberg, Julia Rice, Marco Pistoia, and Ivano Tavernelli.
\newblock Quantum orbital-optimized unitary coupled cluster methods in the strongly correlated regime: Can quantum algorithms outperform their classical equivalents?
\newblock {\em The Journal of Chemical Physics}, 152:124107, 2020.

\bibitem{Tan2021qubitefficient}
Benjamin Tan, Marc-Antoine Lemonde, Supanut Thanasilp, Jirawat Tangpanitanon, and Dimitris~G. Angelakis.
\newblock Qubit-efficient encoding schemes for binary optimisation problems.
\newblock {\em {Quantum}}, 5:454, May 2021.

\bibitem{Verchere2023}
Zoe Verchere, Sourour Elloumi, and Andrea Simonetto.
\newblock { Optimizing Variational Circuits for Higher-Order Binary Optimization }.
\newblock In {\em 2023 IEEE International Conference on Quantum Computing and Engineering (QCE)}, pages 19--25, Los Alamitos, CA, USA, September 2023. IEEE Computer Society.

\bibitem{Visvalingam_Whyatt_1993}
M.~Visvalingam and J.~D. Whyatt.
\newblock Line generalisation by repeated elimination of points.
\newblock {\em The Cartographic Journal}, 30(1):46--51, 1993.

\bibitem{Wang_2020}
Zhihui Wang, Nicholas~C. Rubin, Jason~M. Dominy, and Eleanor~G. Rieffel.
\newblock $\texttt{XY}$-mixers: Analytical and numerical results for the quantum alternating operator ansatz.
\newblock {\em Physical Review A}, 101(1), January 2020.

\bibitem{Zhu_2017_ICCV}
Jun-Yan Zhu, Taesung Park, Phillip Isola, and Alexei~A. Efros.
\newblock Unpaired image-to-image translation using cycle-consistent adversarial networks.
\newblock In {\em Proceedings of the IEEE International Conference on Computer Vision (ICCV)}, Oct 2017.

\bibitem{Zoufal2019}
Christa Zoufal, Aurélien Lucchi, and Stefan Woerner.
\newblock Quantum generative adversarial networks for learning and loading random distributions.
\newblock {\em npj Quantum Information}, 5(1):103, 2019.

\end{thebibliography}

\appendix

\section{Pseudo-code for RDP Algorithm} \label{sec:rdp_pseudo_code}

The pseudo code for the RDP algorithm is shown in Alg.~\ref{alg:DouglasPeucker}. This implementation is based on \cite{rdp_js}, which is the JavaScript equivalent of the recursive implementation of Method 2 of \cite{Douglas_Peucker_1973} using ALGOL W. In the following, the procedure is described in terms of recursive implementation.

First, the first point $S$ and the last point $E$ on the path are taken and recorded. Find the point $P$ on the path where the vertical distance between $P$ and the line defined by $S$ and $E$ is the largest. If this distance is less than or equal to the maximum permissible distance $\epsilon$, the line segment with $S$ and $E$ as endpoints is considered suitable to represent the entire path. If this condition is not satisfied, $P$ is recorded. Next, a subpath \#1 with $S$ as the first point and $P$ as the last point and a subpath \#2 with $P$ as the first point and $E$ as the last point are taken. The same process is performed for each of the two new subpaths, and each of the four new subpaths is further examined... The recursive process ends when the criterion for the maximum permissible distance $\epsilon$ is met. The set of points recorded represents the final compressed path.
\begin{algorithm*}[!ht]
\caption{RDP Algorithm}\label{alg:DouglasPeucker}
\KwIn{PointList $P = \{P_0, P_1, \dots, P_{n-1}\}$, threshold $\epsilon$}
\KwOut{Simplified PointList}
\SetKwFunction{RDP}{RDP}
\SetKwFunction{PerpendicularDistance}{PerpendicularDistance}
\SetKwFunction{Line}{Line}
\SetKwProg{Fn}{Function}{:}{}
\Fn{\RDP{$P$, $\epsilon$}}{
    $d_{max} \gets 0$\;
    $index \gets -1$\;
    $n \gets \text{length}(P)$\;
    \For{$i \gets 1$ \KwTo $n - 1$}{
        $d \gets \PerpendicularDistance(P_i, \Line(P_0, P_{n-1}))$\;
        \If{$d > d_{max}$}{
            $d_{max} \gets d$\;
            $index \gets i$\;
        }
    }
    $ResultList \gets []$\;
    \If{$d_{max} > \epsilon$}{
        $RecResults_1 \gets \RDP(P[0:index+1], \epsilon)$\;
        $RecResults_2 \gets \RDP(P[index:n], \epsilon)$\;
        $n^\prime \gets \text{length}(RecResults_1)$\;
        $ResultList \gets RecResults_1[0:n^\prime - 1] + RecResults_2$\;
        \Return $ResultList$\;
    }
    $ResultList \gets [P_0, P_{n-1}]$\;
    \Return $ResultList$\;
}
\end{algorithm*}

\section{Pseudo-code for the proposed method} \label{sec:pseudo_code}

The pseudo code of the proposed method is shown in Alg.~\ref{alg:our_method}.

First, enumerate the possible forward-connecting edge combinations by traversing each vertex on the path in turn. The determination of connectable edges is done by the process corresponding to Sec.~\ref{sec:edge_thinning}. After the edges are extracted, a combinatorial optimization problem is solved from the viewpoint of 1) the sequence of edges is connected from the start point to the end point, and 2) the number of edges is minimized. This combinatorial optimization problem is formulated using the HOBO formula. The HOBO formula is then converted to the Ising Hamiltonian, and the optimal combination is obtained using QAOA. Finally, the optimal parameters are used for sampling to determine the selected edges. The vertices at both ends of the edge are extracted, sorted, and returned.
\begin{algorithm*}[!ht]
\caption{The proposed method’s algorithm} \label{alg:our_method}
\KwIn{PointList $P = \{p_0, p_1, \ldots, p_{n-1}\}$, threshold $\epsilon$, repeat $n\_reps$}
\KwOut{Decimated points list}
\SetKwFunction{ProposedMethod}{ProposedMethod}
\SetKwFunction{isValidEdge}{is\_valid\_edge}
\SetKwFunction{appendList}{append\_list}
\SetKwFunction{defineHobo}{define\_hobo}
\SetKwFunction{getIsing}{get\_ising}
\SetKwFunction{getHamiltonian}{get\_hamiltonian}
\SetKwFunction{createQaoaAnsatz}{create\_qaoa\_ansatz}
\SetKwFunction{qaoa}{qaoa}
\SetKwFunction{setParams}{set\_params}
\SetKwFunction{calcSelectedEdges}{calc\_selected\_edges}
\SetKwFunction{addSet}{add\_set}
\SetKwFunction{sorted}{sorted}
\SetKwProg{Fn}{Function}{:}{}
\Fn{\ProposedMethod{$P$, $\epsilon$}}{
    $edge\_list \gets []$\;
    \For{$i \gets 0$ \KwTo $n-1$}{
        \For{$j \gets i+1$ \KwTo $n-2$}{
            \If{\isValidEdge{$P, i, j, \epsilon$}}{
                \appendList{$edge\_list, (i, j)$}\;
            }
        }
    }
    $hobo \gets$ \defineHobo{$P, edge\_list$}\;
    $ising \gets$ \getIsing{$hobo$}\;
    $observables \gets$ \getHamiltonian{$ising$}\;
    $qaoa\_ansatz \gets$ \createQaoaAnsatz{$ising, n\_reps$}\;
    $opt\_params \gets$ \qaoa{$qaoa\_ansatz, observables$}\;
    $opt\_ansatz \gets$ \setParams{$qaoa\_ansatz, opt\_params$}\;
    $selected\_edges \gets$ \calcSelectedEdges{$opt\_ansatz, edge\_list$}\;
    $decimated\_points \gets \{\}$\;
    \ForEach{$edge \in selected\_edges$}{
        \addSet{$decimated\_points, edge[0]$}\;
        \addSet{$decimated\_points, edge[1]$}\;
    }
    \Return \sorted{$decimated\_points$}\;
}
\end{algorithm*}

In Alg.~\ref{alg:our_method}, \texttt{IS\_VALID\_EDGE} is the process corresponding to Sec.~\ref{sec:edge_thinning} and is implemented as in Alg.~\ref{alg:is_valid_edge}.

In a sequence of points $P$ representing a path, the edges between the base vertex (indexed by idx) and N vertices ahead are verified. For each edge, it determines whether the vertical distance between the midway vertices (sandwiched between both ends) and the edge are less than or equal to the maximum permissible distance $\epsilon$, and returns true if this condition is satisfied for all edges. Otherwise, false is returned.
\begin{algorithm*}[!ht]
\caption{Edge Validation Algorithm} \label{alg:is_valid_edge}
\KwIn{PointList $P = \{p_0, p_1, \ldots, p_{n-1}\}$, index $idx$, offset $forward$, threshold $\epsilon$}
\KwOut{Boolean indicating if the edge is valid}
\SetKwFunction{isValidEdge}{is\_valid\_edge}
\SetKwFunction{perpendicularDistance}{perpendicular\_distance}
\SetKwFunction{Line}{Line}
\SetKwProg{Fn}{Function}{:}{}
\Fn{\isValidEdge{$P$, $idx$, $forward$, $\epsilon$}}{
    $line \gets$ \Line{$P[idx], P[forward]$}\;
    \For{$i \gets idx + 1$ \KwTo $forward - 1$}{
        \If{\perpendicularDistance{$P[i], line$} $> \epsilon$}{
            \Return False\;
        }
    }
    \Return True\;
}
\end{algorithm*}

\end{document}